\documentclass[10pt]{iopart}

\usepackage[plainpages=false,breaklinks=true]{hyperref}
\usepackage{graphicx,color}
\usepackage{tabularx}
\usepackage{iopams}

\usepackage{framed}
\usepackage{float}

\newcommand{\cbr}{C_{\mathrm{bro}}}
\newcommand{\calv}{C_{\mathrm{A}}}
\newcommand{\cart}{C_{\mathrm{a}}}
\newcommand{\cven}{C_{\bar{\mathrm{v}}}}

\newcommand{\cinh}{C_{\mathrm{I}}}
\newcommand{\cmeas}{C_{\mathrm{measured}}}

\newcommand{\hen}{\lambda_{\mathrm{b:air}}}
\newcommand{\lmuca}{\lambda_{\mathrm{muc:air}}}
\newcommand{\lmucb}{\lambda_{\mathrm{muc:b}}}
\newcommand{\lmamb}{\frac{\lmuca}{\lmucb}}
\newcommand{\lmbma}{\frac{\lmucb}{\lmuca}}

\newcommand{\qbr}{q}

\newcommand{\qalv}{\dot{V}_{\mathrm{A}}}
\newcommand{\qc}{\dot{Q}_{\mathrm{c}}}

\newcommand{\pr}{k_\mathrm{pr}}
\newcommand{\met}{k_{\mathrm{met}}}

\newcommand{\vbro}{\tilde{V}_{\mathrm{bro}}}
\newcommand{\valv}{\tilde{V}_{\mathrm{A}}}

\newcommand{\vh}{\mathbf{h}}
\newcommand{\vb}{\mathbf{b}}
\newcommand{\vc}{\mathbf{c}}

\newcommand{\di}{\mathrm{d}}
\newcommand{\R}{\mathbb{R}}

\newcommand{\lk}{\left(}
\newcommand{\rk}{\right)}

\begin{document}

\title[Modeling-based determination of physiological parameters of systemic VOCs]{Modeling-based determination of physiological parameters of systemic VOCs by breath gas analysis, part 2 }

 \author{Clemens Ager$^{1,2}$,  
    Karl Unterkofler$^{1,5}$, 	
 	Pawel Mochalski$^1$,
 	Susanne Teschl$^3$,
 	Gerald Teschl$^4$,
	Chris~A.\ Mayhew$^{1}$, 
	and
	 Julian King$^1$
	}
\address{$^1$Breath Research Institute, University of Innsbruck, Rathausplatz 4, A-6850 Dornbirn, Austria}
\address{$^2$Univ.-Clinic for Anesthesia and Intensive Care, Innsbruck Medical University, Anichstr. 35, A-6020 Innsbruck, Austria}
\address{$^{3}$University of Applied Sciences Technikum Wien,  H\"ochst\"adtplatz 6, A-1200 Wien, Austria}
\address{$^{4}$Faculty of Mathematics, University of Vienna, Oskar-Morgenstern-Platz 1, 1090 Wien, Austria}
\address{$^{5}$University of Applied Sciences Vorarlberg, Hochschulstr.\ 1, A-6850 Dornbirn, Austria}

%\ead{gerald.teschl@univie.ac.at}
\ead{karl.unterkofler@fhv.at}

\begin{abstract}
In a recent paper \cite{unterkofler2015} we presented  a simple two compartment model which describes 
the influence of inhaled concentrations on   exhaled  breath concentrations for  volatile organic compounds (VOCs) with small Henry constants.
 
In this paper we extend this investigation  concerning the influence of inhaled concentrations on   exhaled  breath concentrations for   VOCs with higher Henry constants.

To this  end we extend our model with an additional compartment which takes into account the influence of the upper airways 
on exhaled breath VOC concentrations.
 
\end{abstract}
%\pacs{ \MSC 92C45, 92C35 }
%\keywords
Version: 15 March 2018\\
J. Breath Res. {\bf 12}, 036011 (2018)

\maketitle

\renewcommand{\thefootnote}{\arabic{footnote}}

  \section{Introduction} 
  In  their paper \cite{spanel2013} \v{S}pan\v{e}l et al.\  investigated the short-term effect of inhaled volatile organic compounds (VOCs) on  exhaled breath concentrations.
They showed for seven different VOCs with very different Henry constants (blood:air partition coefficients) that the exhaled breath concentration closely resembles an affine function (straight line) of the inhaled concentration.

This  motivated our theoretical investigation \cite{unterkofler2015}  regarding the impact of inhaled concentrations for VOCs with low blood:air partition coefficients, i.e., compounds with exhalation kinetics that are described by the Farhi equation \cite{Farhi1967e}. 
For these VOCs the exhaled end-tidal breath concentration resembles the alveolar concentration.

Here we extend this investigation to VOCs with higher  blood:air partition coefficients where the influence of the upper airways cannot be neglected. 
For such VOCs the exhaled end-tidal breath concentration does not equal the alveolar concentration but the bronchial concentration.

Consider for example acetone with typical concentrations of 1 [$\mu$g/l] in breath.  
Assuming that the exhaled end-tidal breath concentration equals  the alveolar concentration and 
using the Farhi equation\footnote{The Farhi equation  \cite{Farhi1967e} relates
the mixed venous concentration $\cven$ with the alveolar concentration $\calv$ by
\begin{eqnarray}
\calv = \frac{\cven}{\hen + r}. \nonumber
\end{eqnarray}
Here $\hen$ is the blood:air partition coefficient and $r$   is the ventilation-perfusion ratio which is approximately $1$ at rest. }
the blood:air partition coefficient {  (dimensionless Henry constant)}  of acetone {  $\hen \approx 340$ (from table 2 in \cite{anderson2003}) } would lead to a  concentration of 0.341 [mg/l] in blood 
which differs considerably from  typically measured  values 
in blood of 1 [mg/l]. 

 Hence one can not neglect the influence of the upper airways when investigating VOCs with higher partition coefficients, see e.g., \cite{anderson2003}. 
    
  \section{A three compartment model} \label{section2}

\begin{figure}%[H]
\centering
\begin{picture}(7,10)

\put(0.7,9.8){$C_\mathrm{I}$}
\put(0.8,9.7){\vector(0,-1){0.85}}
\put(1.4,9.8){$C_\mathrm{bro}$}
\put(1.5,8.85){\vector(0,1){0.85}}
\put(1,9.2){$\qalv$}

\put(3.9,8.15){\vector(-1,0){0.6}}
\put(3.9,8.15){\line(0,-1){3.8}}
\put(4.8,8.1){\parbox{1cm}{bronchial\\compartment}}
\put(4,8){\rotatebox{270}{$\qbr  \dot{Q}_\mathrm{c}$}}
\put(0.5,8){\fbox{\rule[-3ex]{0pt}{8ex}\parbox{1cm}{\vspace{-1mm}\centering $C_\mathrm{bro}$\\[2mm]$V_\mathrm{bro}$} \quad \parbox{1cm}{\vspace{-1mm}\centering $C_\mathrm{muc}$\\[2mm]$V_\mathrm{muc}$} }}
\multiput(1.8,8.7)(0,-0.2){6}{\line(0,-1){0.1}}
\put(1.3,7.05){$D$}
\put(1.15,7.45){\vector(0,-1){0.6}}
\put(1.15,6.85){\vector(0,1){0.6}}

\put(2.5,7.45){\line(0,-1){0.6}}
\put(2.5,7.125){\vector(1,0){1.4}}
\put(4.8,6.1){\parbox{1cm}{alveolar\\compartment}}
\put(4,6.5){\rotatebox{270}{$(1-\qbr) \dot{Q}_\mathrm{c}$}}
\put(0.5,6){\fbox{\rule[-3ex]{0pt}{8ex}\parbox{1cm}{\vspace{-1mm}\centering $C_\mathrm{A}$\\[2mm]$V_\mathrm{A}$} \quad \parbox{1cm}{\vspace{-1mm}\centering $C_\mathrm{c'}$\\[2mm]$V_\mathrm{c'}$} }}
\multiput(1.8,6.7)(0,-0.2){6}{\line(0,-1){0.1}}

\put(4.8,3.65){\parbox{1cm}{body\\compartment}}
\put(0.5,3.65){\fbox{\rule[-9ex]{0pt}{16ex}\parbox{16mm}{\vspace{1mm}\centering $C_\mathrm{body\ blood }$\\[2mm]$V_\mathrm{body \ blood}$\\[4mm] $C_\mathrm{body \ tissue}$\\[2mm]$V_\mathrm{body \ tissue}$} }}
\multiput(0.6,3.75)(0.2,0){9}{\line(1,0){0.1}}

\put(3.9,4.35){\vector(-1,0){1.45}}
\put(3.45,3.3){$k_\mathrm{met}$}
\put(2.45,3.4){\vector(1,0){0.95}}
\put(3.45,2.7){$k_\mathrm{pr}$}
\put(3.4,2.8){\vector(-1,0){0.95}}
\put(0.25,4.35){\line(0,1){0.9}}
\put(0.5,4.35){\vector(-1,0){0.25}}
\put(0.25,5.25){\line(1,0){2.25}}
\put(2.5,5.25){\vector(0,1){0.2}}

\end{picture}
\vspace{-20mm}
\caption{Sketch of the model structure. The body is divided into three distinct functional units: bronchial/mucosal compartment (gas exchange), alveolar/end-capillary compartment (gas exchange) and body compartment (metabolism and production). Dashed boundaries indicate a diffusion equilibrium. Thus in each case  two compartments can be combined into one compartment with an effective volume $\tilde V$, 
e.g., the body blood compartment and the body tissue compartment are assumed to be in an equilibrium and therefore can  
be combined into one single body compartment with an effective volume,
$\tilde V_{B}= V_\mathrm{body \ blood}+ \lambda_{B:b} V_\mathrm{body \  tissue}$.
For more details about effective volumes compare appendix A.2 in \cite{king2010a}. 
The conductance parameter $D$ has units of volume divided by time and quantifies an effective diffusion barrier between the bronchial and the alveolar tract.}\label{fig:model_struct}
\end{figure}
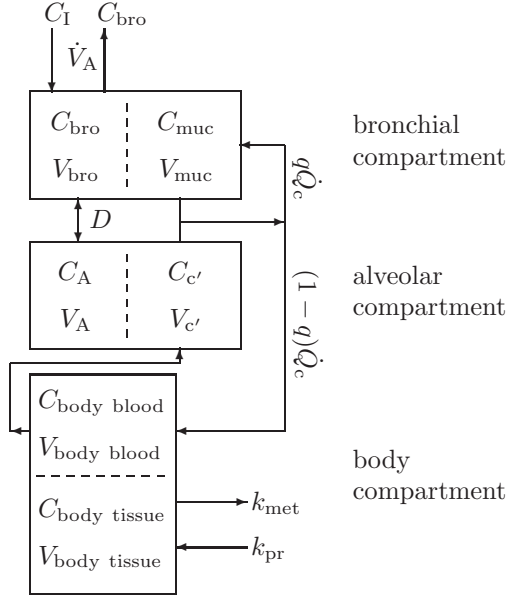

To incorporate the influence of the upper airways on exhaled VOC concentrations we choose the simplest possible model.
It consists of three compartments as sketched in Figure~\ref{fig:model_struct}: a two compartment lung (bronchioles and alveoli) as used in \cite{king2010a} 
and one body compartment.

We consider the bronchial compartment separated into a gas phase and a mucus membrane, which is assumed to inherit the physical properties of water and acts as a reservoir. The part of a VOC dissolved in this layer is transferred to the bronchial circulation, whereby the major fraction of the associated venous drainage is postulated to join the pulmonary veins via the post capillary anastomoses~\cite{lumbbook}.

The amount of a VOC transported  at time $t$  via exhalation and inhalation to the bronchial compartment 
equals therefore
\begin{eqnarray}
\qalv\big(\cinh-\cbr\big), \nonumber
\end{eqnarray}
where $\qalv$ denotes the ventilation,  $\cinh$ denotes the concentration in the inhaled air (normally assumed to be zero), 
and $\cbr$ the bronchial air  concentration\footnote{Note: we have suppressed the time variable $t$, i.e., we 
write $\qalv$ instead of $\qalv(t)$, and so on.}. Moreover, we state that the measured (exhaled) end-tidal breath concentration equals the bronchial level, i.e.,
\begin{eqnarray}\label{eq:meas}
\cmeas=\cbr. \nonumber
\end{eqnarray}

 The contribution of the blood flow through the pulmonary veins via the post capillary anastomoses is
 \begin{eqnarray}  
  \qbr\, \qc\lk\cart-\frac{\lmuca}{\lmucb}\cbr \rk, \nonumber
\end{eqnarray}
where $q$ denotes the fractional blood flow through the bronchioles, $\qc$ the cardiac output, $\cart$ the arterial blood concentration,
 $\lmucb$ the mucus:blood partition coefficient, and  $\lmuca$ the temperature dependent mucus:air partition coefficient (see Appendix~B for details).
 
Then  the arterial blood concentration $\cart$ is given by
\begin{eqnarray}\label{eq:cart}\cart =(1-\qbr)\hen\calv+\qbr\lmamb \,\cbr   %\nonumber
\end{eqnarray}
with $\hen$ denoting the blood:air partition coefficient and $\calv$ the alveolar concentration.
 
 The exchange between the bronchial compartment and the alveolar compartment is modeled as a diffusion process
\begin{eqnarray}  
D\, (\calv-\cbr)\nonumber
\end{eqnarray}
with a diffusion constant $D$ which takes values between zero and infinity.

Thus the total mass balance for the bronchial compartment  reads
\begin{eqnarray} \label{eq:bro}
\hspace{-20mm}
\vbro \frac{\di\cbr}{\di t} = \qalv(\cinh-\cbr)+D(\calv-\cbr) +\qbr\, \qc\lk\cart-\frac{\lmuca}{\lmucb}\cbr \rk. 
\end{eqnarray}

Analogously we derive  the mass balance equations from Figure~\ref{fig:model_struct} for the alveolar compartment  
\begin{eqnarray}\label{eq:alv}
\hspace{-20mm} \valv \frac{\di\calv}{\di t}=D(\cbr-\calv) +(1-\qbr)\qc\big(\cven -\hen\calv\big), 
\end{eqnarray} 
and the body compartment
\begin{eqnarray}\label{eq:body}
\hspace{-20mm} \tilde V_{B} \frac{ d C_{B} }{d t} &=  (1-\qbr)\dot{Q}_{c} (  C_{a}-\cven) -    \met \lambda_{b:B}  C_{B}+ \pr \label{eq1b}, 
\end{eqnarray}   
where $\met$ denotes the total metabolic rate\footnote{We assume that the ambient air is not severely contaminated and hence  metabolism can be modeled with a linear kinetics. }
 of the body and $\pr$ the production rate.
$\vbro$, $\valv$, and $\tilde V_{B}$ denote the effective volume of the bronchiols, alveoli, and the body, 
respectively. $C_{B}$ is the concentration 
in the body which is connected to the mixed venous concentration $\cven$ by Henry's law $\cven = \lambda_{b:B} \, C_{B}$
where $\lambda_{b:B}$  denotes the blood:body tissue partition coefficient.

Remark:  A single body compartment  can be derived from the combination of the liver and tissue compartment of the model in \cite{king2010a}.

Thus the three compartment model for VOCs with higher Henry constant consists 
of the system of the  three linear differential equations (\ref{eq:bro}) -- (\ref{eq:body})
\begin{eqnarray} \label{eq:system}
\hspace{-20mm}\vbro \frac{\di\cbr}{\di t} &= \qalv(\cinh-\cbr)+D(\calv-\cbr) +\qbr\qc\lk\cart-\frac{\lmuca}{\lmucb}\cbr\rk, \nonumber \\
\hspace{-20mm} \valv \frac{\di\calv}{\di t}& =D(\cbr-\calv) +(1-\qbr)\qc\big(\cven -\hen\calv\big), \nonumber\\
\hspace{-20mm}\tilde V_{B} \frac{ d C_{B} }{d t} &=  (1-\qbr)\dot{Q}_{c} (  C_{a}-\cven) - \met  \cven+ \pr .
\end{eqnarray}

Remarks: (i) Summing up these three linear differential equations yields the total change of mass $m_{tot}$ of a VOC, i.e., 
\begin{eqnarray} 
\hspace{-20mm}
\vbro \frac{\di\cbr}{\di t}+\valv \frac{\di\calv}{\di t}+\tilde V_{B} \frac{ d C_{B} }{d t} = \frac{ d m_{tot} }{d t}=
\qalv \cinh-\qalv \cbr+\pr-\met  \cven . \label{eq:totalmass}
\end{eqnarray}
Equation (\ref{eq:totalmass}) shows that the total change of mass of a VOC is given by what is inhaled  minus what is exhaled plus what is produced by the body minus what is eliminated by metabolism (metabolism includes all losses, e.g., by liver,  urine, skin, etc.),  so that the total mass balance is fulfilled.

(ii) In general,  ventilation $\qalv$ and cardiac output $\qc$ are non-constant functions of time. Nevertheless one can show that all solutions of the system~(\ref{eq:system}) starting in $\R_{> 0}^{3}$ remain bounded (see appendix~B, proposition~1 in \cite{king2010a}).

(iii) Rearranging Equation~(\ref{eq:system}) % by the effective volumes 
yields a system of  the form
\begin{eqnarray}  
\frac{\di\vc(t)}{\di t}= N \vc(t) + \vh %\tilde \vb
\end{eqnarray} 
for the vector  $\vc$ of the three  concentrations $\left( \cbr, \calv, \cven \right)$, i.e.,
\begin{eqnarray} 
 \vc = \left( c_1, c_2, c_3 \right)=\left( \cbr, \calv, \cven \right). \nonumber
\end{eqnarray} 
If ventilation $\qalv$ and cardiac output $\qc$ are kept constant and assuming that the production $\pr$ is constant, too, the solution of this system can be given explicitly 
(see, e.g., chapter~3.2 in \cite{teschlode}\footnote{A pdf version of this book is available from\\  \url{http://www.mat.univie.ac.at/~gerald/ftp/book-ode/index.html}}).
All eigenvalues of the constant matrix $N$ are negative and the concentrations approach exponentially 
(the eigenvalues of $N$ are the exponential constants) the equilibrium state $\vc(\infty) =-N^{-1}  \vh$.

When in a stationary state, namely where all quantities and concentrations are constant, the left hand sides of the system
(\ref{eq:system}) are zero and the system of differential equations 
%(\ref{eq:system})
 reduces to a linear algebraic system of the form
\begin{eqnarray}  \label{stationarysystem}
 M \vc = \vb
\end{eqnarray} 
where the matrix $M$ and the vector $\vb$ are given by 
\begin{eqnarray}
\hspace{-20mm}M=
\left( \begin{array}{ccc}
 \qalv + D + q (1-q) \lmamb \qc  & -D - q (1-q) \hen\,  \qc  & 0 \\
-D &  D + q (1-q) \hen\,  \qc   & -   (1-q)   \qc\\
- q (1-q) \lmamb \qc &   -   (1-q)^2 \hen\,  \qc  &\met +  (1-q)   \qc
\end{array} \right)  \nonumber ,
\end{eqnarray}

\begin{eqnarray}
\vb =
\left( \begin{array}{c}
 \qalv \, \cinh \\
0 \\
 \pr 
\end{array} \right) .  
\end{eqnarray}

Trivial linear algebra lets us  write the solution of the system~(\ref{stationarysystem}) with the help of Cramer's rule

\begin{eqnarray} \label{solutions}
\hspace{-20mm} \cbr = c_1 = \frac{\det (M_1)}{\det (M)},\quad \calv = c_2 = \frac{\det (M_2)}{\det (M)},
\quad \cven = c_3 = \frac{\det (M_3)}{\det (M)}
\end{eqnarray}
where $M_j$ denotes the matrix $M$ where the $j$-th column, $j=1,2,3$, is replaced by the vector $\vb$
and $\det(M)$ denotes the determinant of a matrix $M$.

From equation~(\ref{solutions})  we conclude 
% that if all parameters such as $\qalv $, etc.\ are kept constant 
that all concentrations are indeed affine functions (straight lines) of the inhaled concentration $\cinh$.
$\cinh$ appears in the first component of the vector $\vb$ only. Hence $\det(M)$ is independent of $\cinh$. 
The multilinearity of the determinant of the matrix $M_j$  implies the affine dependence on $\cinh$, i.e.,
 \begin{eqnarray} \label{eq:affin}
 c_j(\cinh) = a_j \, \cinh + b_j  ,
  \end{eqnarray}
  where $a_j $ and $b_j, j=1,2,3$ are dependent on $D, \qalv$, etc.
  
For the special case $D=0$ (this is the case for very high partition coefficients 
$\hen >100$)\footnote{The decoupled case $D=\qbr=0$ will be excluded from now on
 as it lacks physiological relevance.}
 we  get  $  \calv = 1/\hen \, \cven$ and furthermore  
 \begin{eqnarray} \label{eqn7}
\cbr(\cinh) & =& a_1 \,\cinh + b_1 , \nonumber\\
\calv(\cinh) &=& a_2 \, \cinh + b_2  , \nonumber\\
\cven(\cinh) &=& a_3 \, \cinh + b_3
 \end{eqnarray}  
 with   
 \begin{eqnarray} \label{eqn8}
a_1 & = &  \frac{1}{1+ \frac{\lmuca}{\lmucb}\, \frac{\qc}{\qalv} \frac{ \qbr (1-\qbr) }{1+\qbr (1-\qbr)\frac{\qc}{   \met}}},
\nonumber\\
b_1 & = &  \cbr(0) = \frac{\pr}{\qalv + \met (\frac{\lmuca}{\lmucb}\, + \frac{\qalv}{\qc} \frac{1}{\qbr (1-\qbr)}) }\nonumber, \\
a_2 & = & \frac{a_3}{\hen}, \quad b_2 = \calv(0)  =   \frac{b_3}{\hen}, \nonumber\\
a_3 & = & \frac{1}{\frac{\lmucb}{\lmuca}\, + \met(\frac{1}{\qalv}+ \frac{1}{\qc} \frac{1}{q(1-q)} \frac{\lmucb}{\lmuca}) }  ,
\nonumber \\
b_3 & = &  \cven(0) = \frac{\pr}{  \met + \frac{\qalv}{ \frac{\lmuca}{\lmucb}\, + \frac{\qalv}{\qc} \frac{1}{\qbr (1-\qbr)} }}.
 \end{eqnarray}  
 Furthermore, the connection between the mixed venous blood concentration and the measured exhaled concentration is given by 
 \begin{eqnarray} 
\cven(0) &=&  \big(\frac{\lmuca}{\lmucb} +\frac{1}{q(1-q)} \frac{\qalv}{\qc} \big) \, \cbr(0). %\nonumber
\end{eqnarray}  
For exogenous VOCs (i.e., $\pr=0$) we have $b_1=b_2=b_3=0$ and $ \cbr(\cinh)   =  a_1 \,\cinh$,\ $\cven(\cinh)  =  a_3 \, \cinh $ which yields
 \begin{eqnarray} \label{eq:exogenous}
 \cven(\cinh) &=& \frac{\frac{\lmuca}{\lmucb}}{1+\frac{1}{q(1-q)} \frac{\met}{\qc}}\, \cbr(\cinh).
  \end{eqnarray}   
Since the fractional blood flow of  the bronchial circulation $q$ is very small ($\qbr \approx 0.01$ \cite{lumbbook}) we have $\qbr (1-\qbr) \approx \qbr$ and the following approximations
are valid 
\begin{eqnarray}
a_1 &     \approx & \frac{1}{1+ \frac{\lmuca}{\lmucb}\, \frac{\qc}{\qalv} \frac{ \qbr   }{1+\qbr  \frac{\qc}{   \met}}} 
\nonumber,\\
b_1  & \approx & \frac{\pr}{\qalv + \met (\frac{\lmuca}{\lmucb}\, + \frac{\qalv}{\qc} \frac{1}{\qbr }) } 
\nonumber ,\\
a_3 &  \approx &\frac{1}{\frac{\lmucb}{\lmuca}\, + \met(\frac{1}{\qalv}+ \frac{1}{\qc} \frac{1}{q} \frac{\lmucb}{\lmuca}) }   
,\nonumber 
\\
   b_3 &   
   \approx & \frac{\pr}{  \met + \frac{\qalv}{ \frac{\lmuca}{\lmucb}\, + \frac{\qalv}{\qc} \frac{1}{\qbr } }} .
    %\nonumber  
 \end{eqnarray}  
Further simplifications are possible  under further assumptions,
e.g., $  \met \to 0$ leads to $a_1=1$ or   $  \met  \approx \qc$ leads to $b_3 \approx  \frac{\pr}{  \met }$.\\

Remarks: (i) Looking at the equation $\cbr(\cinh)  = a_1 \,\cinh + b_1 $ we see that  $ b_1 $ is the contribution to the exhaled breath  by the endogenous production when no room concentration is present and $(1-a_1)$ is the proportion of the room concentration which is taken up by the body.\\

(ii) For $D\not =0$ the calculation is straight forward but the expressions are quite lengthy.
However, these calculation can be easily done with a computer algebra system, e.g., using {\it Mathematica}.
The results are supplied in \ref{AppE}.\\ % online.\\

\subsection{Correction method in order to account for inhaled VOC concentrations} \label{subtract} 

From Equation~(\ref{eq:affin}) we conclude that to correct the measured exhaled concentration for the inhaled one, one has simply to subtract the inhaled concentration multiplied by the gradient $a_1$, i.e.,
  \begin{eqnarray}
 C_{\mathrm{exhaled}}(0)= \cbr(0) = b_1  =\cbr(\cinh)- a_1 \cinh. \label{eqnCorr}
  \end{eqnarray} 
Example 1: With the data from Section~\ref{data} we therefore get for acetone
 \begin{eqnarray}
\cbr(0)= \cbr(\cinh)- 0.384\, \cinh=\cmeas- 0.384\, \cinh.
 \end{eqnarray}
Example 2:
To  estimate  $a_1$ for ethanol we use the following  nominal values: 
$ q=0.01$, $\qalv=5.2$~[l/min], $\qc=6$~[l/min] (from table~1 and 2 in \cite{king2010a}), 
$\met=0.15$~[l/min] ($=$ 7~[g/h] from \cite{cederbaum2012}), $\hen= 1756$ (from \cite{jones1983b}), $\lmuca=2876.7$ at 32$^\circ$ C, $\lmucb=1.17$ (from \cite{staudinger2001}). 
This yields
 \begin{eqnarray}
\cbr(0)= \cbr(\cinh)- 0.047\, \cinh=\cmeas- 0.047\, \cinh.
 \end{eqnarray}
This shows that in contrast to methane \cite{Szabo2016} where one must subtract the total inhaled concentration, for ethanol the inhaled concentration is nearly 
neglectable.

\subsection{Endogenous production and metabolic rates} \label{rates}

The question remains how to determine the endogenous production rate and the total metabolic rate of the body using the theoretical framework introduced above?
When in a stationary state, the averaged values of ventilation and perfusion are constant, then
 Equation~(\ref{eq:affin})  resembles an affine function (straight line) of the form

  \begin{eqnarray}
\cbr (\cinh) &  =  a_1 \, \cinh + b_1, \label{eq17}
 \end{eqnarray}
 $\cinh$ being the variable here.
The constants $a_1$ and $b_1$ are given for $D=0$ by Equation~(\ref{eqn8}).

However, for all cases of $D$ the constants $a_j$ and $b_j,\  j=1,2,3$  are completely  determined by the physiological  
quantities $ \dot{V}_{A}, \dot{Q}_{c},
\pr, \met, q$, and  partition coefficients. The gradient $a_1$ is independent of $\pr$,  fulfills $0<a_1\leq1$, and  depends on
the metabolic rate $\met$ but not the production rate $\pr$.
 The quantity $b_1= \cbr(0)$ is proportional to the production rate $\pr$.

Varying $\cinh$, one can measure $\cbr(\cinh)$   experimentally and thus determine $a_1$ and $b_1$.
Measuring in addition ventilation and  perfusion 
 allows for calculating the total production rate and the total metabolic rate of the body from these two equations.
 For $D=0$ this yields
\begin{eqnarray}  
  \met &=  \frac{q(1-q)(1-a_{1})\, \qc }{(1+\frac{\lmuca}{\lmucb}\, q(1-q)\, \frac{\qc}{\qalv})\, a_{1} -1} , \label{kmet}\\
  \pr &= \frac{b_{1} \, \qc}{a_{1}\frac{\qc}{\qalv} +\frac{\lmucb}{\lmuca} \frac{1}{q(1-q)}(a_{1}-1)}, \label{kprod1}
  \end{eqnarray}
  or
\begin{eqnarray}    
  \pr &= b_{1} \big(\qalv + \met (\frac{\lmuca}{\lmucb}\, + \frac{\qalv}{\qc} \frac{1}{\qbr (1-\qbr)}) \big)  \label{kprod}
 \end{eqnarray}
if $\met$ is already known.

 Remarks: (i)     Note that the numerators in Equations (\ref{kmet}) and (\ref{kprod1}) are small which will cause large errors when there are no good data available.  

(ii) For $D\not =0$ the calculation is straightforward, too, but the expressions are also quite lengthy.
The results are supplied in \ref{AppE}. %as a supplement. \\ % online.\\

\subsection{Test of the theory with  data available from literature} \label{data}

Since \v{S}pan\v{e}l et al.\ did not provide any data for blood flow (cardiac output $\qc$) and breath flow
(alveolar ventilation $\qalv$)  we took the data for {\em acetone} provided by  Wigaeus \cite{wigaeus1981} (i.e., series 1). 
 This data which we have already used in \cite{king2010a} are listed in Table~\ref{table:data}. Note that  $D$  equals zero at rest for acetone.
\begin{table}[H]  
\centering 
\caption{List of data and determined parameters values from \cite{wigaeus1981} and \cite{king2010a}.}  \label{table:data} 
\vspace{1mm}
\begin{tabular}{|lcc|}\hline
 {\large\strut} Parameter & Symbol & value\\ \hline \hline
{\large\strut} inhaled air concentration & $\cinh$ & 1.309 [mg/l]\\
{\large\strut}  exhaled concentration &  $C_{\mathrm{exhaled}}$ &  0.504 [mg/l]\\
 {\large\strut} Diffusion & $D$ & 0 [l/min]\\
 {\large\strut} alveolar ventilation & $\qalv$ &6 [l/min]\\
{\large\strut}	cardiac output & $\qc$ & 5.8 [l/min] \\
{\large\strut} fractional bronchial blood flow & $\qbr$ & 0.0043\\
{\large\strut} blood:air partition coefficient &  $\hen$ & 340\\
{\large\strut} mucus:air partition coefficient (32$^\circ$ C)&  $\lmuca$ & 498\\
{\large\strut} mucus:blood partition coefficient (37$^\circ$ C) &  $\lmucb$ & 1.15\\
{\large\strut} mean bronichal concentration  &  $\cbr(0)$ & 0.0016 [mg/l]\\
\hline
\end{tabular}
\end{table}
This data determine $a_1=0.384$ {($\approx  {C_\mathrm{exhaled}}/\cinh$ for $\cinh > > \cbr(0)$)} and $ b_1=0.0016$ in Equation~(\ref{eqn8}).
 Then the following values can be calculated from  Equation~(\ref{eqn8}). They  are listed in Table~\ref{table:values}.

\begin{table}[H]  
\centering 
\caption{List of calculated values}  \label{table:values} 
\vspace{1mm}
\begin{tabular}{|lccc|}\hline
 {\large\strut} Parameter & Symbol & value& {value in \cite{king2010a}}\\ \hline \hline
{\large\strut} metabolic rate & $\met$& 0.21 [l/min] & 0.18 [l/min]\\
{\large\strut} production rate &  $\pr$ & 0.24 [mg/min]& 0.19 [mg/min]\\
 {\large\strut}	 mixed venous concentration & $\cven(0)$ &1.079 [mg/l]& 1.0 [mg/l] \\
 {\large\strut} alveolar air  concentration &  $\calv(0)$ &0.0032 [mg/l] &0.0029 [mg/l]\\
{\large\strut}	 arterial concentration & $\cart(0)$ & 1.077 [mg/l]& 0.98 [mg/l]\\
\hline
\end{tabular}
\end{table}
These values are in good agreement with the values from the more detailed model developed in \cite{king2010a}.

%\newpage

\section{Discussion}
%{\color{red}   FIXME:  correct} \\

  In  this paper we extended our investigation of  the short-term effect\footnote{This is the typical situation in a clinical examination.} of inhaled volatile organic compounds (VOCs) 
  on  exhaled breath concentrations to VOCs with higher Henry constants. For such VOCs the exhaled end-tidal breath concentration does not equal the alveolar concentration but equals the bronchial concentration and hence it is essential to take  the influence of the upper airways into account.
  
In particular, a special focus is given to the case when the inhaled (e.g., ambient air) concentration is significantly different from zero.
The model elucidates a novel approach for computing metabolic/production rates of systemic VOCs with high blood:air partition coefficients
from the respective breath concentrations. Moreover, it clarifies how breath concentration of such VOCS should be corrected 
{(see Equation~\ref{eqnCorr})} when the inhaled concentration cannot be neglected. 
The model  predicts  an affine relationship (straight line) between exhaled breath concentrations and inhaled concentrations as  shown by
measurements by Spanel et al.\ \cite{spanel2013}
and are in good agreement with data available from Wigaeus \cite{wigaeus1981}. 

The gradient of this line is  
completely  determined by the physiological  
quantities $ \dot{V}_{A}, \dot{Q}_{c},
\pr, \met, q$, and  partition coefficients. However, for practical use it might be easier to determine this gradient directly by experiments for the VOC one is interested in. Note that the gradient {$a_1$ is approximately $  {C_\mathrm{exhaled}}/\cinh$ if $\cinh > > \cbr(0)$}.
Even labeled\footnote{{$^{13}$C labeling is preferred to avoid  D-H-exchanges (article in preparation) when labeling with D-atoms.} }
inhaled VOCs might be used to exclude effects from endogenous production.

Nevertheless, a number of limitations should be mentioned here. Firstly, in order to apply this model for the estimation of metabolic/production rates,  further studies with a representative number of patients will be necessary. In particular, the individual and population ranges of these quantities will have to be determined.
In addition, it should be investigated how these parameters vary with age, body mass, sex, etc..
To circumvent the intricate measurements of ventilation and perfusion, one could measure heart frequency and breath frequency 
and deduce ventilation and perfusion from these parameters.  

In order to account for long-term exposure, the model should be extended to incorporate a storage compartment which fills up and depletes according to its partition coefficient. 
 This yields then at least a 4-compartment model.  
However, for short-term exposure experiments the influence of such a storage compartment will merely be reflected by a slightly different metabolic rate.

\ack%{Acknowledgment}
J.K., P.M.,  and K.U.\  gratefully acknowledge support from the Austrian Science Fund (FWF) under Grant No.\ P24736-B23.
P.M.\ also acknowledges financial support from the Austrian Research Promotion Agency (FFG) for 
the program KIRAS Security Research under the grant DHS-AS.
Furthermore this work has also received funding from the European Union's Horizon 2020 research and innovation 
program under grant agreement no.~644031. We thank the government of Vorarlberg (Austria) for its generous support.

\setcounter{footnote}{0}
 
\appendix

\section{List of symbols} \label{appA} % {\color{red}   FIXME:  } check if list is complete
Table \ref{table:param} summarizes the list of symbols used in the text.
\begin{table}[H]
\centering 
\caption{ Abbreviations}  \label{table:param} 
\vspace{2mm}
\begin{tabular}{|lc|}\hline
 {\large\strut} Parameter & Symbol \\ \hline \hline
{\large\strut}	cardiac output & $\qc$   \\
{\large\strut} alveolar ventilation & $\qalv$ \\
{\large\strut}  ventilation-perfusion ratio &  $r=\qalv/\qc$ \\
{\large\strut} effective volume of    alveoli &$\valv$ \\
{\large\strut}  effective volume of  the body\ &  $\tilde V_{B}$ \\
{\large\strut}  effective volume of  the bronchioles \ &  $\vbro$ \\
{\large\strut} inhaled air concentration & $\cinh$ \\
{\large\strut}	 bronchial concentration & $\cbr$ \\
{\large\strut}	 arterial concentration & $\cart$ \\
{\large\strut} alveolar air  concentration &  $\calv$ \\
{\large\strut}	averaged mixed venous concentration & $\cven$  \\
{\large\strut}  exhaled (measured) concentration &  $C_{\mathrm{exhaled}}=\cmeas$  \\
{\large\strut} body concentration &  $C_{B}$  \\
{\large\strut} metabolic rate & $\met$   \\
{\large\strut} production rate &  $\pr$ \\
{\large\strut} blood:air partition coefficient &  $\hen$ \\
{\large\strut} blood:body partition coefficient &  $\lambda_{b:B} $ \\
{\large\strut} mucus:blood partition coefficient &   $\lmucb$\\
{\large\strut} mucus:air partition coefficient  &  $\lmuca$  \\
{\large\strut} fractional blood flow through  bronchioles &  $q$  \\
\hline
\end{tabular}
\end{table}

\section{Temperature dependence of $\lmuca$  {($=\lambda_{water:air}$)}  }

There is strong experimental evidence that airway temperature constitutes a major determinant for the pulmonary exchange of highly soluble VOCs, cf.~\cite{jones1982}. How this influences the $\lmuca(T)$ partition coefficient was described in detail for acetone in \cite{king2010a}. However, this  can immediately be adapted to other highly soluble VOCs.

The decrease of  solubility in the mucosa -- expressed as the water:air partition coefficient $\lmuca$ -- with increasing temperature can be described in the ambient temperature range by a van't Hoff-type equation~\cite{staudinger2001}
\begin{equation}
\label{eq:hentemp}\log_{10}\lmuca(T)=-A+\frac{B}{T+273.15},
\end{equation}
where $A$ and $B$ (in Kelvin) are proportional to the entropy and enthalpy of volatilization, respectively. 

$\hen$ will always refer to 37$^{\circ}$C. 
Similarly, the partition coefficient between mucosa and blood $\lmucb$ is treated as a constant defined by
\begin{equation}\label{eq:lmucb}
\lmucb:=\lmuca(37^{\circ}\mathrm{C})/\hen. 
\end{equation} 
 Note, that if the airway temperature is below 37$\phantom{}^{\circ}$C we always have that
\begin{equation}
\label{eq:thermdis}
\lmuca/\lmucb \geq \hen.
\end{equation}
as $\lmuca$ is monotonically decreasing with increasing temperature.
In a typical situation the absolute sample humidity at the mouth is 4.7\% 
(corresponding to a temperature of $T \approx 32^{\circ}\mathrm{C}$ 
and ambient pressure at sea level, cf.~\cite{mcfadden1985,hanna1986}). 
Thus the local solubility of a VOC in the mucus layer increases considerably from  the lower respiratory tract up to the mouth, thereby predicting a drastic reduction of air stream VOC concentrations along the airways. 

 { Remark: A comprehensive compilation of water:air partition coefficients including their temperature dependence is given in \cite{sander2015}. Moreover, this reference also discusses the various forms of units used for Henry constants in different fields and the corresponding conversion factors.}
 
\section{Estimation of the  blood-air partition coefficient}

The blood-air partition coefficient can be estimated using the method of Poulin \& Krishnan \cite{Poulin1995}
\begin{equation} 
\hen=\lambda_{o:w}\lambda_{w:a} ( a + 0.3\, b)+ \lambda_{w:a}(c+ 0.7\, b) \label{eq:lwa}
\end{equation}
where, $a=0.0033$ is the fraction of neutral lipids in blood, $b=0.0024$ is the fraction of phospholipids in blood, $c=0.82$ is the fraction of water in blood,  $\lambda_{o:w}$   is the octanol:water partition coefficient and $\lambda_{w:a}$   is the  water:air partition coefficient. 
Equation~(\ref{eq:lwa}) shows the close correlation between $\hen$ and $\lambda_{w:a} =: \lmuca$.

\section{Converting breath VOC concentrations to different conditions}

 When we measure a room concentration of a breath VOC, we measure the temperature $t$ [C], 
the air pressure $p$ [kPa], the relative humidity $h_{ r}$ [\%], and the VOC concentration $C_{ room}$ in, e.g., 
parts per billion [ppb]$=$ [nmol/mol]\footnote{The advantage of [ppb] is that is independent of $p$ and $V$.}.

Since we use conservation laws for modeling we have to convert relative concentrations into [mol/l] (counting number of particles) or
[g/l] (mass balance).

To convert relative concentrations into [mol/l] we must divide this concentration by the volume of one mole $V_m$.
The volume of one mole  can be calculated using the ideal gas law which is sufficiently accurate for trace gases
 \begin{eqnarray}  
  p\ V = n\ R\ T .\nonumber
\end{eqnarray}
Here $n$ denotes the number of moles, $R=8.3144598$\footnote{see \url{http://physics.nist.gov/cgi-bin/cuu/Value?r}} the gas constant, and $T=(273.15+t)$ the absolute temperature. Hence 
as can be seen from
 \begin{eqnarray}  
 V_m = \frac{R\ T}{p} \nonumber
\end{eqnarray}
the volume of one mole depends on pressure and temperature.

To convert relative concentrations further into [g/l] we must in addition multiply with the molar mass $m_m$ of the VOC.

In addition we have to take into account the humidity of the room air.
Humidity is the amount of water in gas form in air. It can be measured as relative humidity $h_r$ (unit  [\%])  
defined as  ratio of the partial pressure of water vapor $p_{H_2O}(t)$ (absolute humidity) to the 
equilibrium vapor pressure of water  $p^{*}_{H_2O}(t)$ at a given temperature 
\begin{eqnarray}  
 h_r = 100\ \frac{p_{H_2O}(t)}{p^{*}_{H_2O}(t)}.\nonumber
\end{eqnarray}

The vapor equilibrium pressure of water  is the pressure at which water vapor is in thermodynamic equilibrium with its condensed state. It depends solely on the temperature $t$ and can be computed  accurately enough 
by the Buck equation\footnote{\url{https://en.wikipedia.org/wiki/Arden_Buck_equation}} 

 \begin{eqnarray} 
p^{*}_{H_2O}(t)= 0.61121 \exp \left( \Big(18.678 -  \frac{t}{234.5} \Big) \Big(\frac{t}{257.14 + t }\Big)\right).
\nonumber
\end{eqnarray}
Here $t$ is measured in [C] and $p$ in [kPa]. 

Thus the fractional pressure $f_{p,w}$ of the absolute humidity is given by    
 \begin{eqnarray} 
f_{p,w}(t,h_r)=  \frac{p_{H_2O}(t)}{p} = \frac{1}{100}\ h_r \frac{p^{*}_{H_2O}(t)}{p} .
\nonumber
\end{eqnarray}
This lets us convert  the measured concentration $C_{room}(t)$ of a VOC to dry conditions by
\begin{eqnarray} 
C_{room, dry}(t) =C_{room}(t)\ \frac{1}{1-f_{p,w}(t,h_r)}.
\nonumber
\end{eqnarray}

When we breathe  air into the lungs it is warmed up to  body temperature $t_{body}=37$ [C] and moisturized to 100\% humidity. 
However, the pressure is immediately balanced.
Using the ideal gas equation for constant pressure we arrive at

\begin{eqnarray} 
C_{lung, dry}(t_{body})= C_{room, dry}(t) \ \frac{T}{T_{body}}.
\nonumber
\end{eqnarray}
In addition when we take  100\% humidity into account we end up with

\hspace{-5mm}
\begin{eqnarray} 
 &C_{lung}(t_{body})= C_{lung, dry}(t_{body}) (1-f_{p,w}(t_{body},100))\nonumber \\ 
= &C_{room}(t)\ \frac{(273.15+t)}{(273.15+t_{body})} \frac{(p-p^{*}_{H_2O}(t_{body}))}{(p- p^{*}_{H_2O}(t)\ h_r/100)}.
\nonumber
\end{eqnarray}

Examples: For $t=22$ [C] the influence of the temperature on the concentration is about 5\%.
\begin{eqnarray} 
\frac{T_{22}}{T_{body}}=\frac{295.15}{310.15} = 0.95.
\nonumber
\end{eqnarray}
For a pressure of $p=100$ [kPa] and a relative humidity of $50$ \% the influence of moistening on the concentration is also about 5\%.
\begin{eqnarray} 
\frac{(p-p^{*}_{H_2O}(37))}{(p- 0.5\ p^{*}_{H_2O}(22)  )}=\frac{(100-6.27988)}{(100-1.3221 )}= 0.95.
\nonumber
\end{eqnarray}
Together this gives a correction factor of about $0.9$.

What we denote by $\cinh$ is hence $C_{lung}(t_{body})$, which is $C_{room}(t)$ converted to body conditions.\\

\vspace{2mm}
Remark: For $t=34$ [C] we get $\frac{T_{34}}{T_{body}}=\frac{307.15}{310.15} = 0.99$ or for $t=32$ [C] we get $\frac{T_{32}}{T_{body}}=\frac{305.15}{310.15} = 0.98.$\\

\vspace{2mm}
Hence a temperature difference between body or lung  compartment and the bronchial compartment can safely be ignored since there is no measurable effect on concentrations. 

\section{The general case where $D\not = 0$.} \label{AppE}

%{\bf Remark:} This appendix will be provided as an online supplement only.\\

Here  we present the general form of the coefficients $a_j, b_j, j=1,2,3$ where the diffusion constant $D$ is not zero, i.e.,
 \begin{eqnarray} \label{eq:affinA}
 c_j(\cinh) = a_j (D) \, \cinh + b_j  (D),
  \end{eqnarray}
  and $a_j $, $b_j, j=1,2,3$ are dependent on $D, \qalv$, etc. 
  
  In addition we did not 
  introduce dimensionless quantities (e.g., $r:= \frac{\qalv}{\qc}$, etc.) to get a more compact form for these coefficients
  since we did not want to introduce a batch of new symbols.
 However, we rearranged the coefficients in such a way that the limit $D\to 0$ ($\hen > 100$ large enough) or $D\to \infty$ (upper airways have no influence) can  be read off directly.

 \begin{eqnarray} \label{eqn7A}
\cbr(\cinh) & =& a_1(D) \,\cinh + b_1 (D), \nonumber\\
\calv(\cinh) &=& a_2 (D)\, \cinh + b_2 (D) , \nonumber\\
\cven(\cinh) &=& a_3 (D)\, \cinh + b_3(D)
 \end{eqnarray}  
 with

%\hspace{-15mm} 
%\begin{minipage}
\begin{eqnarray} \label{eqn8a}
\hspace{-25mm}    a_1    =  
  \frac{1+ D \left(\frac{1 +(1-q) \frac{\qc}{\met} }{(1-q) \hen \qc(1+(1-q) q \frac{\qc}{\met})}\right) }{1+ \frac{\lmuca}{\lmucb}\, \frac{\qc}{\qalv} \frac{ \qbr (1-\qbr) }{1+\qbr (1-\qbr)\frac{\qc}{   \met}}+D \left( \frac{1+ (1-q) \frac{\qc}{\met} + (1-q)^2\hen \frac{\qc}{\qalv}+ q (1-q)  \lmamb \frac{\qc}{\qalv}}{(1-q) \hen \qc(1+(1-q) q \frac{\qc}{\met}) }\right)}, 
\nonumber\\
\hspace{-25mm}    b_1   =    
 \frac{\pr  \left( 1+\frac{D}{ (1-q) q \hen \qc} \right) }{\qalv + \met (\frac{\lmuca}{\lmucb}\, + \frac{\qalv}{\qc} \frac{1}{\qbr (1-\qbr)}) 
+ D \left( \frac{\qalv}{\qc} \frac{1}{\qbr (1-\qbr)\hen} + \frac{\met}{\qc} \left( \frac{1}{q}+\frac{\lmamb }{(1-q)\hen}  +\frac{\qalv}{ (1-q)^2 q \hen \qc }    \right) \right)  }\nonumber, \\
\hspace{-25mm}    a_2   =  
 \frac{1+ D \left( \frac{\met + (1-q) \qc}{ (1-q)^2q \lmamb \qc^2}\right) }{ 
\hen  \left( \lmbma +\met \left(\frac{1}{\qalv}+ \frac{1}{(1-q) q \lmamb \qc}\right)\right) + D \left( \frac{\met \hen+\frac{\qalv}{(1-q)}+ 
\met \left(\frac{q}{(1-q)} \lmamb+\frac{\qalv}{(1-q)^2 \qc}\right) }{q \lmamb \qc \qalv} \right) }, \nonumber\\
\hspace{-25mm}    b_2  = 
\frac{\pr\left( 1+ D \frac{1}{(1-q) q \lmamb \qc+\qalv}\right)}{ \hen \left(\met+ \frac{\qalv}{\lmamb+\frac{\qalv}{(1-q) q \qc}} \right)
+ D \left( \frac{\qalv+ \met \left( (1-q) \hen+ q \lmamb+ \frac{\qalv}{(1-q) \qalv}\right)}{(1-q) q \lmamb \qc+\qalv}\right)}
\nonumber\\
\hspace{-25mm}  a_3   =  
 \frac{1 +D \frac{1}{\qc} \left( \frac{1}{(1-q)  \hen }+ \frac{1}{q} \lmbma \right) }{\frac{\lmucb}{\lmuca}\, + \met(\frac{1}{\qalv}+ \frac{1}{\qc} \frac{1}{q(1-q)} 
\frac{\lmucb}{\lmuca}) +D \left( \frac{ \qalv+\met \left((1-q) \hen+q \lmamb+\frac{\qalv}{(1-q)  \qc} \right) }{(1-q) q \hen \lmamb \qc \qalv} \right) }  ,
\nonumber \\
 \hspace{-25mm}  b_3   =   
  \frac{\pr \left( 1+ D  \left( \frac{(1-q)+ \frac{q}{\hen} \lmamb + \frac{\qalv}{(1-q) \hen \qc}   }{(1-q) q \lmamb \qc+\qalv} \right)\right)}{
   \met + \frac{\qalv}{ \frac{\lmuca}{\lmucb}\, + \frac{\qalv}{\qc} \frac{1}{\qbr (1-\qbr)} }+D \left( \frac{\frac{\qalv}{\hen}+\met \left( (1-q)+ \frac{q}{\hen}\lmamb +
\frac{\qalv}{(1-q) \hen \qc}\right)}{(1-q) q \lmamb \qc+\qalv}\right)}.
 \end{eqnarray}  
%\end{minipage}
 
Taking the limit $D\to 0$ we immediately recover the results in Equation~(\ref{eqn8}).

Taking the limit $D\to \infty $ and $q\to 0$ we recover the results of  the 2-compartment model of \cite{unterkofler2015}.

  For the metabolic rate and the production rate we get in the general case where $D$ is not zero
\begin{eqnarray}  
 \hspace{-25mm}   \met  = 
  \frac{q(1-q)(1-a_{1})\, \qc + D \frac{(1-a_1)}{\hen}}{\left(1+\frac{\lmuca}{\lmucb}\, q(1-q)\, \frac{\qc}{\qalv}\right)\, a_{1} -1
  +D \left(\frac{a_1 (1-q)}{\qalv}+\frac{q a-1}{\hen \qalv}\lmamb- \frac{1-a_1}{(1-q) \hen \qalv} \right)} ,\nonumber \\ \label{kmetD}\\
 \hspace{-25mm}  \pr = 
 \frac{b_{1}\left(   \qc+ D \left( \frac{1}{\hen (1-q)}+\frac{1}{q} \lmbma\right)\right) }{a_{1}\frac{\qc}{\qalv} -\frac{\lmucb}{\lmuca} \frac{(1-a_{1})}{q(1-q)}+
  D\left(\frac{a_{1}}{a \qalv}\lmbma+\frac{a_1}{(1-q) \hen \qalv}-\frac{1-a_1}{(1-q)^2 q \hen \qalv}\lmbma  \right)}. \nonumber \\  \label{kprod1D}
  \end{eqnarray}

Again taking the limit $D\to 0$ we immediately recover the results in Equation~(\ref{kmet}) and Equation~(\ref{kprod1}).

And taking the limit $D\to \infty $ and $q\to 0$ we recover the results of  the 2-compartment model of \cite{unterkofler2015}.

\section*{References}
\bibliographystyle{amsplain}
\bibliography{AllCit14}

\end{document}